\documentclass[prb,twocolumn,showpacs,superscriptaddress]{revtex4}

\usepackage{graphicx}
\usepackage{amsmath}
\usepackage{amssymb}
\usepackage{epsfig}

\renewcommand{\v}[1]{{\bf #1}}

\newcommand{\w}{{\omega}}

\def\eqa{\begin{eqnarray}}
\def\eea{\end{eqnarray}}
\newcommand{\eq}{\begin{equation}}
\newcommand{\ee}{\end{equation}}
\newcommand{\nn}{\nonumber\\}

\renewcommand{\>}{\rangle}

\newcommand{\p}{\partial}

\newcommand{\ua}{\uparrow}
\newcommand{\da}{\downarrow}
\newcommand{\ra}{\rightarrow}

\newcommand{\al}{\alpha}
\newcommand{\bt}{\beta}
\newcommand{\del}{\delta}

\newcommand{\Ga}{\Gamma}

\newcommand{\La}{\Lambda}

\newcommand{\si}{\sigma}

\usepackage{color}

\begin{document}

\title{Competing electronic orders on Kagome lattices at van Hove filling}

\author{Wan-Sheng Wang}
\affiliation{National Lab of Solid State Microstructures, Nanjing
University, Nanjing, 210093, China}

\author{Zheng-Zhao Li}
\affiliation{National Lab of Solid State Microstructures, Nanjing
University, Nanjing, 210093, China}

\author{Yuan-Yuan Xiang}
\affiliation{National Lab of Solid State Microstructures, Nanjing
University, Nanjing, 210093, China}

\author{Qiang-Hua Wang}
\affiliation{National Lab of Solid State Microstructures, Nanjing
University, Nanjing, 210093, China}

\begin{abstract}
The electronic orders in Hubbard models on a Kagome lattice at van
Hove filling are of intense current interest and debate. We study
this issue using the singular-mode functional renormalization
group theory. We discover a rich variety of electronic
instabilities under short range interactions. With increasing
on-site repulsion $U$, the system develops successively
ferromagnetism, intra unit-cell antiferromagnetism, and charge
bond order. With nearest-neighbor Coulomb interaction $V$ alone
($U=0$), the system develops intra-unit-cell charge density wave
order for small $V$, $s-$wave superconductivity for moderate $V$,
and the charge density wave order appears again for even larger
$V$. With both $U$ and $V$, we also find spin bond order and
chiral $d_{x^2 - y^2} + i d_{xy}$ superconductivity in some
particular regimes of the phase diagram. We find that the $s$-wave
superconductivity is a result of charge density wave fluctuations
and the squared logarithmic divergence in the pairing
susceptibility. On the other hand, the $d$-wave superconductivity
follows from bond order fluctuations that avoid the matrix element
effect. The phase diagram is vastly different from that in
honeycomb lattices because of the geometrical frustration in the
Kagome lattice.

\end{abstract}


\maketitle

\section{Introduction}

The Kagome lattice model has attracted considerable attention duo
to its high degree of geometrical frustration. In the Mott
insulating limit, several possible states have been proposed as
the ground state of the Heisenberg model in this lattice, such as
the $U(1)$ algebraic spin liquid (SL), \cite{WXG} the valance bond
solid, \cite{Huse} the triplet-gapped SL, \cite{WZY} and the
singlet-gapped SL with signatures of $Z_2$ topological order.
\cite{White} On the other hand, several exotic phases have been
proposed for the Kagome Hubbard model, such as the
ferromagnetism at electron filling $1/3$ (or $5/3$) per site,
\cite{Pollmann1} the fractional charge at $1/3$ filling for
spinless fermions,\cite{Pollmann2} and the Mott transition in
anisotropic Kagome lattices.\cite{Furukawa,Yamada}

Of particular interest is the possible phases at the van Hove
filling (the filling fraction is $2/3\pm 1/6$ per site), where the
Fermi surface (FS) is perfectly nested and has saddle points on
the edges of the Brillouine zone. These properties of the normal
state makes it unstable against infinitesimal interactions.
Similar FS appears in triangle and honeycomb lattices and were
shown to develop, under short range repulsive interactions, chiral
spin-density-wave (SDW) state \cite{Martin, LiTao,wws} or chiral
$d_{x^2-y^2} + i d_{xy}$ superconducting
state.\cite{Chubokov,Thomale1} Both states break time-reversal and
parity symmetries, and are topologically nontrivial. Given the
similar FS, a simple FS nesting argument would predict similar
phases in the Kagome model. This seems to be the case in a recent
variational cluster perturbation theory (with an additional spin
disordered phase).\cite{LJX} However, as already realized in
\cite{LJX} and emphasized in \cite{Thomale2}, the interaction
vertex viewed in the band basis has a strong momentum dependence
(matrix element effect). This is because the character of the
Bloch state on the FS depends on the position of the momentum. The
matrix element effect weakens the nesting effect significantly for
a local interaction $U$, leading to a new phase diagram in a
recent analytical renormalization group study.\cite{Thomale2} Such
an analysis would be exact for a featureless fermi surface and
infinitesimal interactions, but its applicability to the case of
finite interactions together with perfect fermi surface nesting
with van Hove singularity is an interesting issue to be addressed.

The functional renormalization group (FRG) method is a
differential perturbation theory with respect to the increment of
the phase space rather than in the interaction itself. It provides
the flow of one-particle irreducible vertex functions versus the
running parameter that controls the phase space.\cite{wetterich}
If implemented exactly the applicability of FRG is not limited by
the size of the interaction. In practice, however, the vertices
are truncated up to the four-point vertices under the assumption
that higher order vertices are irrelevant. The FRG is promising to
address finite interactions and treat particle-particle and
particle-hole channels on equal footing. The applicability of FRG
has been demonstrated in the contexts of cuprates~
\cite{Honerkamp} and iron based superconductors.\cite{WangFa}
Recently, a singular-mode functional renormalization group (SMFRG)
method was developed and applied to investigate topological
superconductivity in correlated electron systems with or near van
Hove singularities.\cite{wws,xyy}

In this paper we perform SMFRG study of the model at van Hove
filling. We discover a rich variety of electronic instabilities
under short range interactions. With increasing on-site repulsion
$U$, the system develops successively ferromagnetism (FM), intra
unit-cell antiferromagnetism (AFM), and charge bond order (CBO).
With nearest-neighbor Coulomb interaction $V$ alone ($U=0$), the
system develops intra-unit-cell charge density wave (CDW) order
for small $V$, $s-$wave superconductivity ($s$SC) for moderate
$V$, and CDW appears again for even larger $V$. With both $U$ and
$V$, we also find spin bond order (SBO) and chiral $d_{x^2 - y^2}
+ i d_{xy}$ superconductivity ($d$SC). Our results are summarized
in the phase diagram Fig.\ref{pd}. We find that the $s$SC is a
result of CDW fluctuations and the squared logarithmic divergence
in the pairing susceptibility. On the other hand, the $d$SC
follows from bond order fluctuations that avoid the matrix element
effect. The phase diagram is vastly different from that in
honeycomb lattices.

The rest of the paper is arranged as follows. In Sec.II, we define
the model and illustrate the matrix element effect. In Sec. III,
we introduce the FRG method. In Sec. IV, we first discuss the
leading instabilities at typical points in the parameter space,
and conclude by a discussion of the phase diagram. Finally, Sec.V
is a summary and perspective of this work.

\section{The model and the matrix element effect}\label{MeanField}

\begin{figure}
\includegraphics[width=8.5cm]{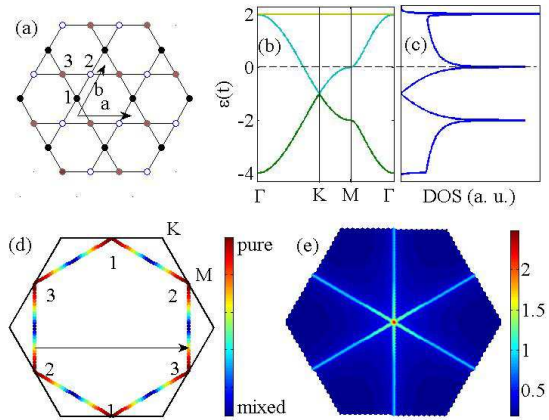}
\caption{(a) Structure of the Kagome lattice. Here $\v a=(1,0)$
and $\v b=(1/2,\sqrt{3}/2)$ are the lattice unit vectors. The
labels $1,2,3$ denote the three sublattices. (b) The tight-binding
dispersion along high symmetric directions. The dashed line is the
Fermi level corresponding to the van Hove filling. (c) Normal
state density of states. (d) Fermi surface and sublattice weights
in the Bloch states thereon. On the endpoint of a fermi surface
segment the Bloch state is contributed completely by one
sublattice as indicated by the numbers. Within the segment the
character are mixed as a superposition of the sublattice
characters on the two endpoints. The arrow indicates one of the
nesting vectors. (e) The largest eigenvalue of $\chi_0(\v q)$ as a
function of $\v q$.} \label{kag}
\end{figure}

The Hubbard model we used for the Kagome lattice is given by \eqa
H = &&-t \sum_{\<ij\>\si} (c^\dagger_{i\si} c_{j\si}+{\rm h.c.}) -
\mu N_e\nn && + U \sum_i n_{i\ua}n_{i\da} + V \sum_{\<ij\>} n_i
n_j, \eea where $t$ is the hopping integral, $\<ij\>$ denotes
bonds connecting nearest-neighbor sites $i$ and $j$, $\si$ is the
spin polarity, $\mu$ is the chemical potential, $N_e$ is the total
electron number operator, $U$ is the on-site Hubbard interaction
and $V$ is the Coulomb interaction on nearest-neighbor bonds.
Fig.\ref{kag}(a) shows the structure of the Kagome lattice. The
different symbols denote the three sublattices, and $\v a$ and $\v
b$ are the two principle translation vectors.  Fig.\ref{kag}(b)
shows the band structure of the model along high symmetry cuts in
the Brillouine zone. The lower two bands cross at the Dirac point.
The highest band is a flat band. The dashed line highlights the
van Hove singularity. Fig.\ref{kag}(c) shows the normal state
density of states. The three sharp peaks arise from the van Hove
singularities in the lower two bands and the third flat band.
Fig.\ref{kag}(d) shows the FS and the character of the
Bloch states thereon. The FS appears to be perfectly nested.
However, the character changes along each segment. The end points
of each segment are saddle points. They have pure but different
sublattice characters. The characters are mixed within the segment
as shown by the color scale.

Consider the local interaction $U$ for the moment. Such an
interaction causes scattering at any wave vector. The nested FS
would favor scattering connected by the nesting vectors and would
lead to antiferromagnetism in usual case. {\em However, since $U$
preserves sublattice indices, the character variation along the FS
causes significant momentum dependence if $U$ is projected to the
band basis}, a matrix element effect as emphasized in
\cite{Thomale2}. This effect hampers the scattering significantly.
To have a better idea of this effect, we calculate the zero
frequency bare spin susceptibility $\chi_0(\v q)$ for site-local
spin densities, where $\v q$ is the momentum transfer. The
susceptibility is a matrix function in terms of the sublattice
labels ($\al$ and $\bt$), \eqa \chi^{\al\bt}_0(\v
q)=-\frac{T}{N}\sum_{\v k,m} G^{\al\bt}(\v k,i\w_m)G^{\bt\al}(\v k
+\v q,i\w_m), \eea where $T$ is the temperature, $N$ the number of
unit cells, $\v k$ the lattice momentum, $\w_m$ the Matsubara
frequency, and $G(\v k, i\w_m)$ the bare Green's function (in the
sublattice basis). Fig.\ref{kag}(e) shows the largest eigenvalue
of $\chi_0(\v q)$ as a function of $\v q$ (for $T=0.001t$).
Instead of isolated peaks we see branch cuts of maxima in the
momentum space. These cuts cross at the origin, where there is in
fact a logarithmic singularity due to the saddle points. (The
singularity is smeared by the finite size and finite temperature
in the calculation). It is clear that site-local ferromagnetism
rather than antiferromagnetism is the most favorable spin order,
in contrast to the case in the honeycomb lattice.\cite{LiTao,wws}
The lesson we learned from the above analysis is that for a
multi-sublattice system the matrix element effect could weaken the
nested scattering and alter the usual intuition regarding FS
nesting. There is, however, a caveat in this kind of Stoner
analysis, since it ignores mode-mode coupling between the
particle-hole channels, and between particle-hole and
particle-particle channels. To treat all channels on equal footing
we now switch to FRG.

\section{The SMFRG method}\label{FRGMethod}

In the following we apply a particular implementation of FRG,
i.e., the SMFRG, which appears advantageous to treat systems with
or near van Hove singularities.\cite{Husemann,wws,xyy} In this
implementation, a generic four-point vertex function $\Ga_{1234}$,
which appears in the interaction $c^{\dag}_1c^{\dag}_2(
-\Ga_{1234})c_3c_4$, where $1=(\v k,\al)$ is a dummy label
indicating the lattice momentum and sublattice label, is
decomposed into the pairing ($P$), the crossing ($C$), and the
direct($D$) channels as \eqa &&\Ga^{\al\bt\mu\nu}_{\v k + \v q,
-\v k, -\v p, \v p + \v q}  \ra
 \sum_{mn} f^*_m(\v k,\al,\bt) P_{mn}(\v q)f_n(\v p,\nu,\mu), \nn
&& \Ga^{\al\bt\mu\nu}_{\v k + \v q, \v p, \v k, \v p + \v q} \ra
 \sum_{mn} f^*_m(\v k,\al,\mu) C_{mn}(\v q)f_n(\v p,\nu,\bt), \nn
&& \Ga^{\al\bt\mu\nu}_{\v k + \v q, \v p, \v p + \v q, \v k} \ra
 \sum_{mn} f^*_m(\v k,\al,\nu) D_{mn}(\v q) f_n(\v
 p,\mu,\bt).\label{decompose}
\eea Here ${f_m}$ is a set of orthonormal lattice form factors. A
form factor defines a particular composite boson with definite
collective momentum in the particle-hole or particle-particle
channel, bearing a definite irreducible representation under the
point group. The fact that the same generic vertex can be
decomposed into different channels reflects the fact that these
channels have mutual overlaps. The momentum space form factors are
related to the real counterparts as, $f_m(\v k,\al,\bt)=\sum_{\v
r\in m} f_m(\v r,\al,\bt)e^{-i\v k\cdot\v r}$ where $\v r$ belongs
to a set of bond vectors connecting sublattices $\al$ and $\bt$
and assigned to $m$. In our practice the bond vectors are
truncated up to those connecting the eighth neighbors (or third
like-sublattice neighbors). In the following we use $m=(l,\al,{\bf
\del})$ to characterize the form factor label $m$, with $l$
indicating the symmetry of the form factor, $\al$ one of the two
sublattice labels, and ${\bf \del}$ a basis bond vector that can
generate the set of bond vectors under the point group. This is
applicable since we set the symmetry center at an atomic site so
that the symmetry group is $C_2$. Under this point group, $\al$
and $\bt$ are invariant. There are only two irreducible
representations $A_g$ (even) and $A_u$ (odd) for $C_2$. We
emphasize that even though the real-space range of the form
factors is truncated the range of composite boson scattering is
unlimited. This enables us to address the thermodynamic limit.

In the SMFRG, $P$, $C$ and $D$ are substituted into independent
sets of one-loop and one-particle irreducible FRG Feynman diagrams
where they would become potentially singular. (For example $P$ is
substituted into the particle-particle diagram.) This leads to the
differential change $\p P$, $\p C$ and $\p D$ with respect to the
change of the running scale $\La$, which we chose as the infrared
cutoff of the Matsubara frequency. Since there are overlaps among
the three channels, the full change is a sum of the partial one
plus the overlaps. It is in this sense that SMFRG takes care of
mode-mode coupling and treats all channels on equal footing. This
enables an initially repulsive pairing channel to become
attractive at low energy scales, and is thus able to reflect the
well-known Kohn-Luttinger anomaly.\cite{L_K} The technical details
have been exposed elsewhere.\cite{wws,note}

The effective interaction in the superconducting (SC), spin
density wave (SDW), and CDW channels are given by $V_{sc} = - P$,
$V_{sdw} = C$, and $V_{cdw} = C - 2D$, respectively. By singular
value decomposition, we determine the leading instability in each
channel, \eq V^{mn}_x(\v q_x) = \sum_\al S^\al_x \phi^\al_x(m)
\psi^\al_x(n),\ee where $x = sc, sdw, cdw$, $S^\al_x$ is the
singular value of the $\al -$th singular mode, $\phi^\al_x$ and
$\psi^\al_x$ are the right and left eigenvectors of $V_x$,
respectively. We fix the phase of the eigenvectors by requiring
$Re[\sum_m \phi^\al_x(m) \psi^\al_x(m)]
> 0$ so that $S^\al_x < 0$ corresponds to an attractive mode in
the $x-$ channel. In the pairing channel $\v q_{sc} = 0 $
addresses the Cooper instability. The ordering wave vector in the
SDW/CDW channel $\v q = \v q_{sdw/cdw}$ is chosen at which
$V_{sdw/cdw}(\v q)$ has the most attractive eigenvalue. We note
that such a vector has symmetry-related images, and may change
during the FRG flow before settling down to fixed values. On the
other hand, given the most singular mode, an effective field can
be defined for the ordered state (or the condensed composite
boson), \eqa &&H_{sc}=\sum_{m,\v k} \psi_{sc}(m)f^*_m(\v
k,\al,\bt)c^\dag_{\v k,\al,\ua}c^\dag_{-\v k,\bt,\da}+{\rm h.c.},
\nn && H_{cdw}=\sum_{m,\si,\v k} \psi_{cdw}(m)f^*_m(\v
k,\al,\bt)c_{\v k+\v q_{cdw},\al,\si}^\dag c_{\v k,\bt,\si}+{\rm
h.c.},\nn && H_{sdw}=\sum_{m,\v k}\psi_{sdw}(m)f^*_m(\v
k,\al,\bt)c^\dag_{\v k+\v q_{sdw},\al,\ua}c_{\v k,\bt,\da}+{\rm
h.c.}, \label{fields} \eea up to global factors. It is understood
that the sublattice labels $\al$ and $\bt$ are determined by $m$
according to our construction of form factors. The order
parameters are encoded in the coefficients in the above field
operators. Two remarks are in order. First there is a residual
SU(2) degeneracy in the case of triplet pairing and in the SDW
order parameters. Second, the order parameters are in general
nonlocal in real space (unless the contributing form factors are
all local).

\section{SMFRG results}\label{FRGresult}

In this section we provide the SMFRG results for the model defined
in the previous section. We begin by discussing the results at
specific points in the parameter space $(U,V)$, and summarize the
systematic results on a dense grid of $(U,V)$ by a phase
diagram.\\

\begin{figure}
\includegraphics[width=8.5cm]{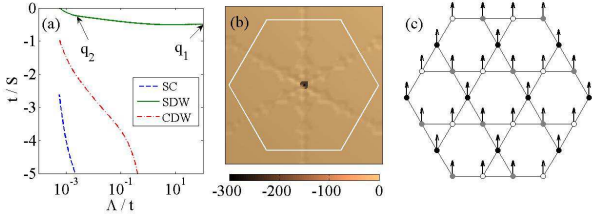}
\caption{(Color online) Results for $U = 2t$ and $V = 0$. (a) FRG
flow of (the inverse of) the most singular values S in the SC
(blue dashed line), SDW( green solid line), and CDW (red
dashed-dot line) channels. (b) The renormalized interaction
$\sum_m V^{mm}_{sdw}$ for $m = (A_g,\al,0)$ ($\al=1,2,3$) as a
function of the collective momentum $\v q$. The hexagon indicates
the Brillouin-zone boundary. (c) The order parameter $h_i$ (drawn
as arrows) associated with the dominant SDW singular mode.}
\label{FM}
\end{figure}

{\em Ferromagnetic order}: For $U = 2t$ and $V = 0 $,
Fig.\ref{FM}(a) shows the flow of the most negative singular
values( denoted as $S$) in the SC, SDW, and CDW channels. Clearly
the SDW (green solid line) is the leading instability. During the
flow $\v q_{sdw}$ evolves from $\v q_1 = (\pi,\pi/\sqrt{3})$ and
settles down at $\v q_2 = 0$. The renormalized interaction $\sum_m
V^{mm}_{sdw}(\v q)$ for $m = (A_{g}, \al, 0)$ ($\al=1,2,3$), which
have dominant value in the leading singular mode, is shown in
Fig.\ref{FM}(b). It has a strong peak at momentum $\v q = 0$.
Because the dominant form factor is local the ordered spin density
is site-local. The effective field operator $H_{sdw}$ according to
Eq.(\ref{fields}) can be rewritten as $H_{sdw}=\sum_{i\si} h_i \si
c^\dag_{i\si}c_{i\si}$ with the order parameter $h_i$ shown in
Fig.\ref{FM}(c). This describes a FM order. The SC and CDW channel
turn out to be sub-leading from Fig.\ref{FM}(a).\\

\begin{figure}
\includegraphics[width=8.5cm]{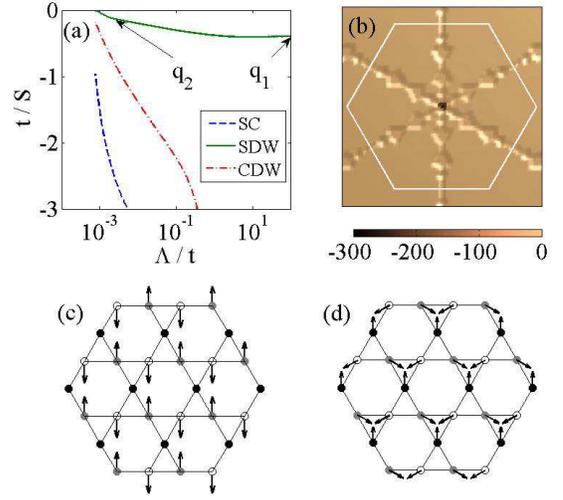}
\caption{(Color online) Results for $U = 2.5t$ and $V = 0$. (a) FRG
flow of the most singular values in the SC (blue dashed line), SDW
(green solid line), and CDW (red dashed-dot line) channels. (b)
The renormalized interaction $\sum_m V^{mm}_{sdw}$ for $m =
(A_g,\al,0)$ ($\al=1,2,3$) as a function of $\v q$. (c) The order
parameter $h_i$ associated with one of the two degenerate SDW
modes. (d) The spin structure in the mean field ordered state
which combines the two degenerate singular modes.} \label{AFM}
\end{figure}

{\em Intra-unit-cell antiferromagnetic order}: For $U=2.5t$ and
$V=0$, the flow of the singular values is shown in
Fig.\ref{AFM}(a). Again the SDW channel is the leading
instability. During the flow, $\v q_{sdw}$ evolves from $\v q_1 =
(\pi, \pi/\sqrt{3})$ and settles down at $\v q_2 = 0$, in the same
fashion as above. Fig.\ref{AFM}(b) shows the interaction $\sum_m
V^{mm}_{sdw}(\v q)$ for $m = (A_{g}, \al, 0)$ ($\al=1,2,3$). It
also has a strong peak at momentum $\v q = 0$. There are in fact
two degenerate singular modes (apart from the SU(2) degeneracy).
One of them leads to the order parameter $h_i$ shown in
Fig.\ref{AFM}(c), with the ratio $0:-1:1$ on the three
sublattices. The other mode lead to a ratio $2:-1:-1$ (not shown).
Both modes are antiferromagnetic within the unit cell, but is
ferromagnetic from cell to cell. Comparing to the FM state, we
call such a state the AFM state, although the ordering momentum is
zero. A mean field analysis shows that in the ordered state the
two degenerate modes are mixed in such a way that the spin patten
is as shown in Fig.\ref{AFM}(d), with an angle of $120^o$ between
nearby spins. The SC and CDW
channels remain to be sub-leading from Fig.\ref{AFM}(a).\\

\begin{figure}
\includegraphics[width=8.5cm]{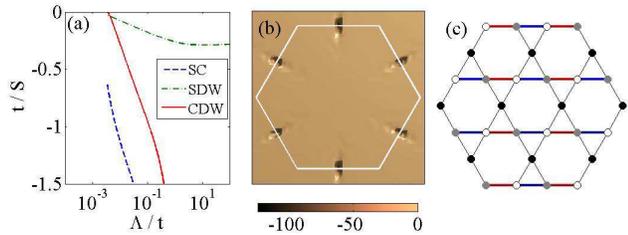}
\caption{(Color online) Results for $U = 3.5t$ and $V = 0$. (a) FRG
flow of the most singular values $S$ in the SC (blue dashed line),
SDW (green dashed-dot line), and CDW (red solid line) channels.
(b) The renormalized interaction $\sum_m V^{mm}_{cdw}$ for $m = (
A_u, 1, 1/4\hat{x}+\sqrt{3}/4\hat{y})$, $m = ( A_u, 2,
1/2\hat{x})$ and $m = (A_u, 3, 1/4\hat{x} -\sqrt{3}/4\hat{y})$, as
a function of $\v q$. The three independent peaks correspond to
the three form factors, respectively. (c) The real space structure
of the order parameter $\chi_{ij}$ associated with one of the
dominant CDW modes with the ordering momentum $\v
Q=(0,2\pi/\sqrt{3})$. The red (blue) color indicates $\chi_{ij}$
is positive (negative).} \label{CBO}
\end{figure}

{\em Charge bond Order}: From Fig.\ref{FM} and Fig.\ref{AFM}, we
find that the CDW channel is enhanced with increasing $U$. This
trend continues until the CDW channel becomes dominant for $U
> 2.85t$. Fig.\ref{CBO}(a) shows the FRG flow for $U = 3.5t$ and $V =
0$. During the flow the $\v q_{cdw}$ evolves but settles down at
$\v q = (0, 2\pi/\sqrt{3})$ (or its symmetric images) in the given
view field. The dominant renormalized interaction $\sum_m
V^{mm}_{cdw}(\v q)$, for $m = (A_{u}, 1,
1/4\hat{x}+\sqrt{3}/4\hat{y})$, $m = (A_{u}, 2, 1/2\hat{x})$ and
$m = (A_{u}, 3, 1/4\hat{x}-\sqrt{3}/4\hat{y})$, is shown in
Fig.\ref{CBO}(b), where we see isolated peaks at the six nesting
vectors (three of which are independent and correspond to the
three form factors). We find that the effective field $H_{cdw}$
constructed according to Eq.(\ref{fields}) for the above singular
modes can be rewritten as
$H_{cdw}=\sum_{\<ij\>\si}\chi_{ij}(c_{i\si}^\dag c_{j\si}+{\rm
h.c.})$, and is thus a CBO state. The pattern of the order
parameter $\chi_{ij}$ depends on the ordering vector $\v Q$. For
$\v Q=\v (0,2\pi/\sqrt{3})$, it is shown in Fig.\ref{CBO} (c).
Notice that the field $\chi_{ij}$ is nonzero on parallel lines
orthogonal to $\v Q$. This is also the case for the other ordering
momenta related to $\v Q$ by $C_{6v}$ operations. Clearly, the CBO
breaks both rotation and translation symmetries. The reason that
the nesting vector is at work here is because the bond-centered
charge density $\sum_\si(c_{i\si}^\dag c_{j\si}+c_{j\si}^\dag
c_{i\si}$) connects different sublattices, and can take advantage
of the inter-saddle scattering connected by the nesting vector.
Notice that this kind of order is already beyond the mean field
theory. It is a result of the overlap between the SDW and CDW
channels as seen from Fig.\ref{CBO}(a) where the SDW channel
dominates at high energy scales. The pairing channel is
still subdominant here.\\

\begin{figure}
\includegraphics[width=8.5cm]{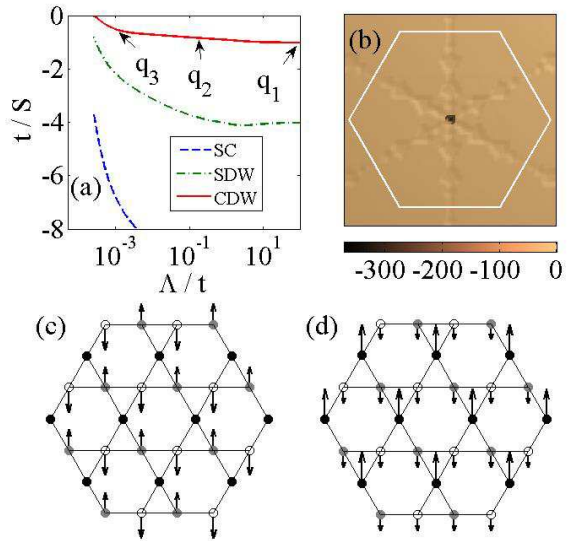}
\caption{(Color online) Results for $U = 0$ and $V = 0.25t$. (a)
The FRG flow the most singular values $S$ in the SC (blue dashed
line), SDW (green dashed-dot line), and CDW (red solid line)
channels. (b) The renormalized interaction $\sum_m V^{mm}_{cdw}$
for $m = ( A_g, \al, 0)$ ($\al=1,2,3$) as function a function of
$\v q$. (c) and (d) show the order parameter $\eta_i$ associated
with the two degenerate CDW singular modes. The length of the
arrows indicate the amplitude and the direction of the arrow
indicate the sign of the order parameter.} \label{CDW}
\end{figure}

{\em Intra-unit-cell Charge density wave}: We now consider the
effect of the nearest neighbor interaction $V$. Fig.\ref{CDW}(a)
shows the FRG flow for $U = 0$ and $V = 0.25t$. It is clear that
the CDW channel (red solid line) is the leading instability.
During the flow the $\v q_{cdw}$ evolves from $\v q_1 = (0,
2/\sqrt{3})\pi$ to $\v q_2 = (0, 0.385)\pi$ and finally settles
down at $\v q_3 = 0$. The dominant renormalized interaction
$\sum_m V^{mm}_{cdw}$ for $m = (A_g, \al, 0)$ ($\al=1,2,3$) shown
in Fig.\ref{CDW}(b) has a sharp peak at $\v q=0$. There are two
degenerate singular modes. The effective field $H_{cdw}$
constructed according to Eq.(\ref{fields}) can be rewritten as
$H_{cdw}=\sum_{i\si}\eta_i c_{i\si}^\dag c_{i\si}$ with the order
parameter $\eta_i$ shown in Fig.\ref{CDW}(c) and (d) for the two
singular modes. This is an intra-unit-cell CDW state. It beaks
rotational symmetry but does not break the translation symmetry.
It is therefore an analogue of the Pomeranchuk instability on
square lattices.\cite{Metzner} The SDW
and SC channels are sub-leading in this case.\\

\begin{figure}
\includegraphics[width=8.5cm]{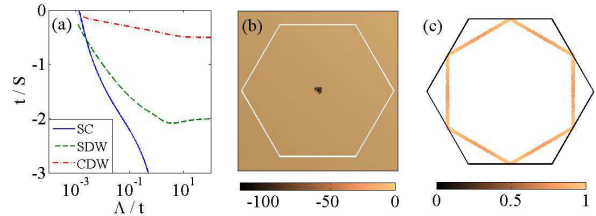}
\caption{(Color online) Results for $U = 0$ and $V = 0.5t$. (a) The
FRG flow the most singular values $S$ in the SC (blue solid line),
SDW (green dashed line), CDW (red dashed-dot line) channels. (b)
The renormalized interaction $V^{mm}_{sc}(\v q)$ for $m = ( A_g ,
1, 0)$ as a function of  $\v q$. (c) The momentum space gap
function on the Fermi surface associated with the SC singular
mode.} \label{sSC}
\end{figure}

{\em $S$-wave superconductivity}: The FRG flow for $U = 0$ and $V
= 0.5t$ is shown in Fig.\ref{sSC}(a). We find that the SC channel
is the leading instability. Fig.\ref{sSC}(b) shows the
renormalized interaction $V^{mm}_{sc}$ for $m = (A_g, 1, 0)$. (By
symmetry, interactions involving form factors centered on the
other sublattices contribute similarly.) Inspection of the
eigenfunction $\phi_{sc}$ reveals that it has dominant values for
$A_g$ form factors involving $\v r=0$ and subdominant values for
$A_g$ form factors involving $\v r$ connecting nearest
like-sublattice neighbors. The gap function from $H_{sc}$
constructed according to Eq.(\ref{fields}) projected on the fermi
surface is shown in Fig.\ref{sSC}(c). Clearly it is an $s$SC gap
function. Such a pairing symmetry persists for small $U>0$.
However the dominant pairing amplitude for $U=0$ is on-site, while
the amplitude on bonds (connecting nearest like-sublattices)
increases and eventually dominates with increasing $U$. Inspection
of Fig.\ref{sSC}(a) reveals that such
an s-wave pairing follows from the overlap with the CDW channel.\\

\begin{figure}
\includegraphics[width=8.5cm]{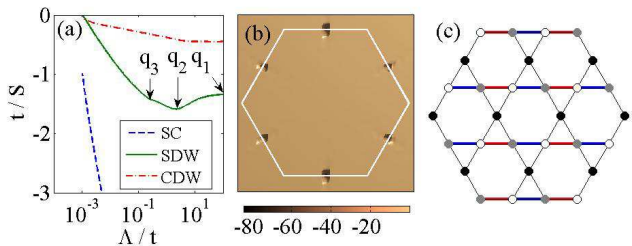}
\caption{(Color online) Results for $U = V = 0.75t$. (a) FRG flow
of the most singular values $S$ in the SC (blue dashed line), SDW(
green solid line), and CDW (red dashed-dot line) channels. (b) The
renormalized interaction $\sum_m V^{mm}_{sdw}$ for $m = ( A_u, 1,
1/4\hat{x}+\sqrt{3}/4\hat{y})$, $m = ( A_u, 2, 1/2\hat{x})$, and
$m = ( A_u, 3, 1/4\hat{x}-\sqrt{3}/4\hat{y})$ as a function of $\v
q$. The three independent peak momenta corresponds to the three
form factors, respectively. (c) The real space structure of the
order parameter $\xi_{ij}$ associated with one of the SDW singular
modes with the ordering momentum $\v Q=(0,2\pi/\sqrt{3})$. The
blue (red) bond indicates that $\xi_{ij}$ is positive (negative).}
\label{SBO}
\end{figure}

{\em Spin bond order}: The FRG flow for $U = V = 0.75t$ is shown
in Fig.\ref{SBO}(a). Clearly the SDW (green solid line) is the
leading instability. During the flow $\v q_{sdw}$ evolves from $\v
q_1 = (0, 2/\sqrt{3})\pi$ to $\v q_2 = 0$ and finally settles down
at $\v q_3 = \v q_1$. Fig.\ref{SBO}(b) shows the renormalized
interaction $\sum_m V^{mm}_{sdw}$ for $m = (A_u, 1,
1/4\hat{x}+\sqrt{3}/4\hat{y})$, $m = (A_u, 2, 1/2\hat{x})$ and $m
= (A_u, 3, 1/4\hat{x}-\sqrt{3}/4\hat{y})$, where we see isolated
peaks at the six nesting vectors (three of which are independent
and correspond to the three form factors). The effective field
operator in the real space can be written as
$H_{sdw}=\sum_{\<ij\>\si}\xi_{ij}\si (c_{i\si}^\dag
c_{j\si}+c_{j\si}^\dag c_{i\si})$ (apart from the SU(2)
degeneracy). The pattern of the order parameter $\xi_{ij}$ depends
on the ordering vector $\v Q$. For $\v Q=\v (0,2\pi/\sqrt{3})$, it
is shown in Fig.\ref{SBO} (c). As in the case of CBO state, the
order parameter $\xi_{ij}$ is nonzero on parallel lines orthogonal
to $\v Q$. This describes an SBO state. The SC and CDW
channel is sub-leading in this case.\\

\begin{figure}
\includegraphics[width=8.5cm]{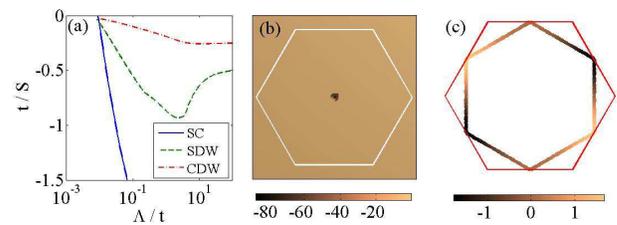}
\caption{(Color online) Results for $U = 2t$ and $V = 1.5t$. (a) FRG
flow the most singular values $S$ in the SC (blue solid line), SDW
(green dashed line), and CDW (red dashed-dot line) channels. (b)
The renormalized interaction $V^{mm}_{sc}(\v q)$ for $m = ( A_g,
1, 1/2\hat{x}+\sqrt{3}/2\hat{y})$ as a function of $\v q$. (c) The
momentum space gap function on the Fermi surface associated with
one of the two degenerate SC singular modes.} \label{dSC}
\end{figure}

{\em Chiral $d_{x^2-y^2} + i d_{xy}$ superconductivity}:
Fig.\ref{dSC}(a) shows the FRG flow for $U = 2t$ and $V = 1.5t$.
Clearly, the SC channel is the leading instability.
Fig.\ref{dSC}(b) shows the renormalized interaction $V^{mm}_{sc}$
for $m = (A_g, 1, 1/2\hat{x}+\sqrt{3}/2\hat{y})$. Such a form
factor shows the pairing is on third-neighbor bonds (or nearest
like-sublattice neighbor bonds). From the singular mode
$\phi_{sc}$ we construct the effective pairing operator $H_{sc}$,
and get the gap function in the momentum space as shown in
Fig.\ref{dSC}(c). This is clearly a $d_{xy}$-wave gap function. In
fact there is another degenerate singular mode which gives a
$d_{x^2-y^2}$-wave gap function (not shown). Using the
renormalized pairing interaction we performed mean field
calculations to find that the ordered state is a chiral $d_{x^2 -
y^2} + i d_{xy}$ superconducting state, which we call the $d$SC
state. The chiral state is fully gapped on the fermi surface and
thus saves more energy. Fig.\ref{dSC}(a) shows that the SDW and
CDW channels are dominant at high energy scales. Inspection of the
later stage of the FRG flow reveals that the singular modes in
these channels contains dominant CBO and SBO components (on
nearest bonds). We shall come
back to this point later.\\

\begin{figure}[h]
\includegraphics[width=8.5cm]{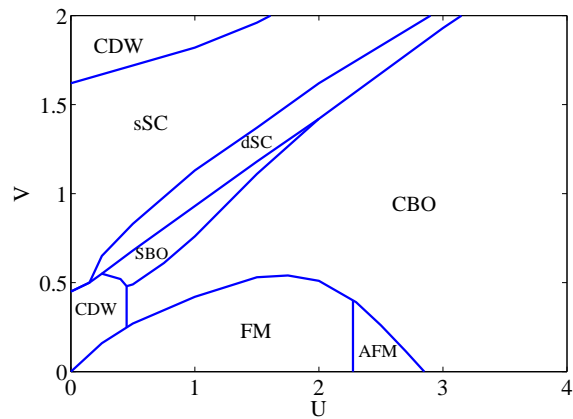}
\caption{The phase diagram of the Kagome lattice at van Hove
filling. The electronic orders and the associated ordering momenta
are: FM ($\v Q=0$), intra-unit-cell AFM ($\v Q=0$), CBO ($\v Q=\v
M$), intra-unit-cell CDW ($\v Q=0$), SBO ($\v Q=\v M$),
$d_{x^2-y^2} + i d_{xy}$-wave SC ($d$SC, $\v Q=0$) and $s$-wave SC
($s$SC, $\v Q=0$). Here $\v M$ is one of the nesting vectors
connecting the saddle points on the fermi surface.} \label{pd}
\end{figure}

{\em The phase diagram}: Apart from the typical results discussed
above,  we have performed systematic SMFRG calculations on a dense
grid in the $(U,V)$ plane. The results are summarized as a phase
diagram shown in Fig.\ref{pd}. The CBO and SBO states have
ordering momenta at one of the nesting vectors, while the others
order at zero momentum without breaking translation symmetry.
However, the CDW and AFM states have intra-unit-cell structures.
This phase diagram can be understood as follows.

Along the $U=0$ axis, the $s$-wave superconductivity appears
inbetween the intra-unit-cell CDW states at small and large values
of $V$. This is counter-intuitive at a first sight since
increasing $V$ would favor CDW further. However, the numerical
result is reasonable for the following reasons. While the CDW
susceptibility behaves as $\ln(W/\La)$ at the running scale $\La$
because of the van Hove singularities in the normal state density
of states (here $W$ is of the order of the bandwidth), the SC
susceptibility diverges as $\ln^2(W/\La)$ due to a further Cooper
instability.\cite{Chubokov} Therefore, once the initially
repulsive pairing channel becomes slightly attractive via the
overlap with the CDW channel, the pairing interaction could grow
in magnitude faster than the CDW interaction, and could eventually
overwhelm the CDW interaction. This explains the emergence of the
$s$-wave superconductivity for moderate $V$. However, if $V$ is
initially small, the overlap with the SC channel is small during
the flow. On the other hand, if $V$ is large enough, the CDW
channel diverges before the SC channel takes advantage of the fast
growth. These considerations are consistent with our results along
the $U=0$ axis.

In the phase diagram we see that both CBO and SBO phases are in
proximity to the $d$-wave SC phase. This is a reasonable result
since we find that the bond orders are on nearest-neighbor bonds,
while the $d$-wave pairing are on third-neighbor bonds (or nearest
like-sublattice neighbor bonds). It is the even order processes
involving the bond-density interactions that have overlap with the
above singlet pairing interaction, which are therefore immune to
the sign structure in the SBO and CBO interactions. On the other
hand, the on-site repulsion disfavors $s$-wave pairing. This makes
$d$-wave pairing viable. Interestingly by utilizing the bond order
fluctuations the pairing mechanism avoids the matrix element
effect that would frustrate site-local spin fluctuations at the
nesting vector.

Along the $V=0$ axis, our SMFRG result predicts the charge bond
order for large $U$. This is indeed a spin disordered phase as
found in \cite{LJX}, and is beyond the mean field theory but
consistent with the lack of a well defined site-local spin ordered
phase. The reason that a large $U$ favors a spin disordered state
rather than local spin moment ordering is twofold. First the
matrix element effect weakens nested scattering and favors
ferromagnetic ordering. Second a sufficiently large $U$ makes the
nested scattering more important as compared to the case of small
$U$. This would favor antiferromagnetic ordering. The site-local
spin ordering is thus frustrated by the competition of
ferromagnetism and antiferromagnetism. The compromise is the CBO
state, which is an analogue of the valence bond solid and reflects
the short-range spin correlations. The limit of $U\gg t$ is beyond
the scope of our SMFRG, but enables mapping of the model to a
doped $t-J$ model. We leave it for further investigations.

Finally, for $U\sim t$ and with increasing $V$, the successive
orders are FM, CBO, SBO, $d$SC, $s$SC and CDW. This sequence is
reasonable as follows. The CBO and SBO states take advantage of
$V$ since by connecting different sublattices it avoids the matrix
element effect. However, a large $V$ favors CDW. In the
intermediate region, the CBO/SBO fluctuations drive $d$SC while
CDW fluctuations drives $s$SC, as discussed above. This explains
why there is a transition from $d$SC to $s$SC with increasing $V$.

\section{Summary and perspective}

In summary, we have studied the extended Hubbard model on Kagome
lattice at van Hove filling using the SMFRG method. We discovered
a variety of phases in the parameter space. Along the $V=0$ axis
and with increasing on-site repulsion $U$, the system develops
successively ferromagnetism, intra unit-cell antiferromagnetism,
and charge bond order. With nearest-neighbor Coulomb interaction
$V$ alone ($U=0$), the system develops intra-unit-cell charge
density wave order for small $V$, $s-$wave superconductivity for
moderate $V$, and CDW appears again for even larger $V$. With both
$U$ and $V$, we also find spin bond order and chiral $d_{x^2 -
y^2} + i d_{xy}$ superconductivity.  We find that the $s$-wave
superconductivity is a result of CDW fluctuations and the squared
logarithmic divergence in the pairing susceptibility. On the other
hand, the $d$-wave superconductivity follows from bond order
fluctuations that avoid the matrix element effect. We summarized
the results by the phase diagram in Fig.\ref{pd}. It is vastly
different from that in honeycomb lattices, and the difference can
be attributed to the frustrating matrix element effect.\\

We notice that the spin $1/2$ Kagome lattice has been realized in
Herbertsmithite $ZnCu_3(OH)_3Cl_2$ \cite{Shores,SHLee} and its
isostructural Mg-based paracatamite
$MgCu_3(OH)_6Cl_2$.\cite{Kermarrec} Also, the optical Kagome
lattice has been simulated experimentally in ultra-cold atomic
gases, and the optical wavelengths can be suitably adjusted for
fermionic isotopes such as $^6Li$ and $^{40}K$.\cite{GBJo} With
the possibility of tuning $U$ and $V$ continuously, the optical
lattice with ultracold atomic gases are most promising to realize
the predictions presented in this paper.\\

During the writing of this paper we became aware of a parallel
work in which a similar scenario is addressed.\cite{ronny} \\

\acknowledgments{QHW thanks Jian-Xin Li and Fan Yang for
interesting discussions. The project was supported by NSFC (under
grant No.10974086 and No.11023002) and the Ministry of Science and
Technology of China (under grant No.2011CBA00108 and
2011CB922101).}


\begin{references}
\bibitem{note} In order to see how the normal state becomes
unstable it is a usual practice to ignore the flow of the
two-point vertex, which is nothing but the inverse of one-particle
Green's function. We also ignore the frequency dependence of the
four-point vertices under the usual assumption that the frequency
dependence is irrelevant.
\end{references}
\end{document}